\documentclass[prd,superscriptaddress,twocolumn,showpacs,showkeys]{revtex4}
\usepackage{amssymb,amsmath,epsfig}

\usepackage{eurosym}
\usepackage{amsfonts}
\usepackage{array}
\usepackage{amsthm}
\usepackage{bm}
\usepackage{palatino}
\usepackage[colorinlistoftodos]{todonotes}
\usepackage{mathpazo}
\usepackage{amssymb}
\usepackage{eurosym}
\usepackage{amsmath}
\usepackage{graphics}
\usepackage{color}
\usepackage{graphicx}
\usepackage{hyperref}
\hypersetup{
    colorlinks=true,
    linkcolor=blue,
    filecolor=magenta, citecolor=blue,    urlcolor=blue,
}

\begin{document}

\title{Weak field deflection angle by regular black holes with cosmic strings using the Gauss-Bonnet theorem}

\author{A. \"{O}vg\"{u}n}
\email{ali.ovgun@pucv.cl}
\homepage{https://www.aovgun.com}
\affiliation{Instituto de F\'{\i}sica, Pontificia Universidad Cat\'olica de Valpara\'{\i}%
so, Casilla 4950, Valpara\'{\i}so, Chile}
\affiliation{Physics Department, Faculty of Arts and Sciences, Eastern Mediterranean
University, Famagusta, North Cyprus, via Mersin 10, Turkey}

\date{\today}

\begin{abstract}
In this paper, we investigate light bending in the spacetime of regular black holes with cosmic strings in weak field limits. To do so, we apply the Gauss-Bonnet theorem to the optical geometry of the black hole; and, using the Gibbons-Werner method, we obtain the deflection angle of light in the weak field limits which shows that the bending of light is a global and topological effect. Afterwards, we demonstrate the effect of a plasma medium on the deflection of light by RBCS. We discuss that increasing cosmic string parameter $\mu$ and mass $M_0$ will increase the bending angle.

\end{abstract}


\maketitle

\section{Introduction}

Black hole physics preserve the mystery of the present, not only by the discovery of gravitational waves \cite{Abbott:2016blz}, but also by the more fundamental level of black hole physics, such as entropy and the information paradox \cite{Mathur:2009hf}. Moreover, in spite of all deep studies on black hole physics, the singularity area of the black hole, where strong curvature effects occur, is unknown and is an open issue in physics until we discover a theory of quantum gravity \cite{Hawking:1976ra}. On the other hand, many different models of black holes are obtained to solve the singularity problem. These non-singular solutions of the black hole, called regular black holes, attracted strong attention in recent years, especially the model with nonlinear electrodynamics coupled to Einstein's theory of gravity. First, Bardeen suggested that regular black holes with magnetic charges obey the weak energy condition \cite{bardeen}. Many various works of Bardeen-like black holes are done using the nonlinear electrodynamics to remove singularities \cite{AyonBeato:2000zs,Hayward:2005gi,Eiroa:2010wm,Toshmatov:2015wga,Sharif:2010pj,Halilsoy:2013iza,Ovgun:2018jbm,Ovgun:2016oit,Kuang:2018goo,Ovgun:2017iwg,Otalora:2018bso,Neves:2014aba,Bayraktar:2018say}. Thus, it is important to check their observational signatures \cite{Akiyama:2019cqa}.

Gravitational lensing is a helpful technique to understand galaxies, dark matter, dark energy and the universe \cite{Bartelmann:2010fz}. Since the first gravitational lensing observation by Eddington, a lot of works on gravitational lensing have been written for black holes, wormholes, cosmic strings and other objects \cite{Keeton:1997by,Bhadra:2003zs,Whisker:2004gq,Chen:2009eu,Nandi:2006ds,Eiroa:2002mk,Mao:1991nt,Bozza:2002zj,Hoekstra:2003pn,Virbhadra:2002ju,Virbhadra:1999nm,Gallo:2011mv,Crisnejo:2017jmx,Sharif:2015qfa,Gibbons:1993cy,Javed:2019qyg}. In 2008, Gibbons and Werner showed a different way to obtain the deflection angle of light from non-rotating asymptotically flat spacetimes \cite{Gibbons:2008rj}, then Werner extended this study to stationary spacetimes \cite{Werner:2012rc}. Their method was based on the Gauss-Bonnet theorem and the optical geometry of the black hole's spacetime, where the source and receiver are located at asymptotic regions. Then Ishihara et al. extended this method for finite-distances (large impact parameter cases) \cite{Ishihara:2016vdc}. Recently, Crisnejo and Gallo have shown that plasma medium deflects photons \cite{Crisnejo:2018uyn}. For more recent works, one can see \cite{Jusufi:2017lsl,Sakalli:2017ewb,Jusufi:2017mav,Ono:2017pie,Jusufi:2017vta,Ovgun:2018prw,Jusufi:2017hed,Arakida:2017hrm,Jusufi:2017vew,Ono:2018ybw,Jusufi:2017uhh,Ovgun:2018xys,Jusufi:2018jof,Ovgun:2018fnk,Ovgun:2018ran,Jusufi:2018kmk,Ovgun:2018oxk,Crisnejo:2018ppm,Ovgun:2018fte,Ovgun:2018tua,Ono:2018jrv}.

The purpose of this work is to study the deflection angle by regular black holes in a topological defect background, given by the cosmic string spacetime \cite{Neves:2014aba} using the Gauss-Bonnet theorem to look at the influence of topological defects \cite{Barriola:1989hx} on gravitational lensing. For comparison, we consider the notion of the deflection angle of massive particles, or the deflection of photons, in a plasma medium from a regular black hole with cosmic strings (RBCS). Our main aim is to demonstrate possible effects of cosmic strings and nonlinear electrodynamics on the deflection angle. 

This paper is composed as follows: in section 2, we briefly review RBCS. In section 3, we calculate the deflection angle by RBCS using the Gauss-Bonnet theorem in weak field regions. Then in section 4, we extend our studies for the deflection of light by RBCS in a plasma medium. We recap our findings in section 5.

\section{Regular black holes with cosmic strings}
\label{sec:1}
The RBCS metric in spherical coordinates is given by the equations \cite{Bayraktar:2018say}
\begin{equation}
d s ^ { 2 } = - f ( r ) d t ^ { 2 } + \frac { d r ^ { 2 } } { f ( r ) } + r ^ { 2 } \left( d \theta ^ { 2 } + \zeta ^ { 2 } \sin ^ { 2 } \theta d \phi ^ { 2 } \right),
\label{RBCS}
\end{equation}

and \begin{equation}f ( r ) = 1 - \frac { 2 m ( r ) } { r }\end{equation}

with the cosmic string parameter $\zeta = 1 - 4 \mu$. It is noted that the mass function \cite{Neves:2014aba} is given by
\begin{equation}
m(r)={\frac{M_0}{{\left[1+{\left(\frac{r_0}{r}\right)}^q\right]}^{\frac{p}{q}}}} , \label{mass}
\end{equation}
where $M_0$ and $r_0$ are mass and length, respectively. 
The above metric reduces as follows for Bardeen black holes ($p=3$, $q=2$) and Hayward black holes ($p=q=3$) \cite{Hayward:2005gi}. There are two solutions for $r_0<M_0$, where $r=r_{\pm}$. Note that the inner horizon is $r_-$ and the outher horizon is $r_+\:{\approx}\: 2m(r)$.

\section{Calculation of deflection angle by RBCS optical spacetime}

The RBCS optical spacetime can be simply written in equatorial plane $\theta=\pi/2$, to obtain null geodesics ($\mathrm{d}s^{2}=0$):
\begin{eqnarray}
\mathrm{d}t^{2}=\frac{\mathrm{d}r^{2}}{f(r)^{2}}+\frac{\zeta^{2}r^{2}\mathrm{d}\varphi^{2}}{f(r)}.
\label{metric2}
\end{eqnarray}
To use Gauss-Bonnet theorem, first, one should obtain the Gaussian curvature $K$-an intrinsic property of optical spacetime. The optical geometry is in two dimensions and is calculated for the RBCS as follows \cite{Gibbons:2008rj}:
\begin{equation}
K=\frac{R_{icci Scalar}}{2} \approx-{\frac {2{\it M_0}}{{r}^{3}}}+{\frac {3{{\it M_0}}^{2}}{{r}^{4}}}.
\label{curvature}
\end{equation}
\begin{itemize}
    \item The Gaussian curvature of the optical RBCS spacetime is negative so that, locally, all the light rays diverge.
    \item There is no contribution from cosmic strings.
    
\item To find multiple images (after converging), one should use theory, such as the Gauss-Bonnet theorem, to connect to local features of spacetime, such as Gaussian curvature.
\end{itemize}

\subsection{The Gauss-Bonnet theorem}

The Gauss-Bonnet theorem is defined for the region $D_ {R} $ in $M$, with boundary $\partial D_{R}=\gamma_{\tilde{g}}\cup C_ {R} $ \cite{Gibbons:2008rj}
\begin{equation}
\iint\limits_{D_{R}}K\,\mathrm{d}S+\oint\limits_{\partial D_{R}}\kappa\,\mathrm{d}t=2\pi\chi(D_{R})-(\theta_{O}+\theta_{S})=\pi.
\label{gaussbonnet}
\end{equation}
\begin{equation}
\iint\limits_{D_{R}}K\,\mathrm{d}S+\oint\limits_{\partial D_{R}}\kappa\,\mathrm{d}t=\pi.
\label{gaussbonnet2}
\end{equation}
Note that the geodesics curvature is given by $\kappa$. For the case of $R$ going to infinity, both jump angles are taken as $\pi/2$,  (shortly $\theta_{O}+\theta_{S}\to \pi$). Since $D_ {R} $ is non-singular, the Euler characteristic is $\chi(D_{R})=1$ and $\kappa(\gamma_{\tilde{g}})=0$. For the near asymptotic limit of $R$, $C_{R}:= r(\varphi)=R=const.$, the radial component of the geodesic curvature is found by:

\begin{equation}
\kappa(C_{R})=|\nabla_{\dot{C}_{R}}\dot{C}_{R}|=\left(\tilde{g}_{rr}\dot{C}^{r}_{R}\dot{C}^{r}_{R}\right)^\frac{1}{2}\to\frac{1}{R}.
\end{equation}
and then
\begin{equation}
\kappa(C_{R})\mathrm{d}t=\frac{1}{R}(\zeta\,R )\,\mathrm{d}\,\varphi.
\end{equation}

 Note that it is not flat because of the cosmic strings at asymptotic limits. The Gauss-Bonnet equation reduces to:
\begin{eqnarray}\label{def1}
\pi=\iint\limits_{S_{\infty}}K\,\mathrm{d}S+\zeta \int\limits_{0}^{\pi+\alpha}\mathrm{d}\varphi.
\end{eqnarray}
where $\alpha$ is a deflection angle and the optical surface area of RBCS is $\mathrm{d}S=\zeta r\mathrm{d}r\,\mathrm{d}\varphi$. 

\subsection{Deflection angle in weak field limits}
In the weak field regions, the light ray follows a straight line approximation, so that we can use the condition of  $r=b/\sin\varphi $ at zeroth-order. After we calculate the optical geometry \eqref{curvature} and use the Gauss-Bonnet theorem \eqref{def1}, the deflection angle is found as follows:
\begin{equation}
\alpha=\frac{\pi-\pi\zeta}{\zeta}-\int\limits_{0}^{\pi}\int\limits_{\frac{b}{\sin \varphi}}^{\infty}K\, r\,\mathrm{d}r\,\mathrm{d}\varphi .
\label{angle}
\end{equation}

The deflection angle $\alpha$ of RBCS in weak field limits is found as follows:
\begin{equation}
\alpha\simeq {\frac {4{\it M_0}}{b}}
+4
\,\pi\,\mu.\label{deflection}
\end{equation}

Note that the cosmic string parameter $\mu$, and the mass term increase the deflection angle. 

\section{Weak gravitational lensing by RBCS in a plasma medium}

In this section, we investigate the effect of a plasma medium on the weak gravitational lensing by RBCS.

The refractive index $n(r)$ for RBCS is obtained as \cite{Crisnejo:2018uyn},
\begin{equation}
n(r)=\sqrt{1-\frac{\omega_{e}^{2}}{\omega^{2}_{\infty}} \bigg(1-\frac{2m(r)}{r}\bigg)},
\end{equation}
where the mass function \cite{Neves:2014aba} is given in \eqref{mass},  $\omega_{e}$ is the electron plasma frequency and $\omega_{\infty}$ is the photon frequency measured by an observer at infinity. Then the corresponding optical metric is:

\begin{equation}
d \sigma ^ { 2 } = g _ { i j } ^ { \mathrm { opt } } d x ^ { i } d x ^ { j } = \frac { n ^ { 2 } ( r ) } { f ( r ) } \left( \frac{d r ^ { 2 }}{f(r)} + \zeta^2 r^2 d \varphi ^ { 2 }\right)
\end{equation}

\begin{widetext}
The Gaussian curvature for the above optical metric is calculated as follows: 
\begin{equation}
\mathcal{K}={\frac {{\it M_0}\, \left( {\omega_{e}}^{2}-2\,{\omega_{\infty}}
^{2} \right) {\omega_{\infty}}^{2}}{ \left( {\omega_{e}}^{2}-{
\omega_{\infty}}^{2} \right) ^{2}{r}^{3}}}-3\,{\frac {{{\it M_0}}^{2}
 \left( {\omega_{e}}^{2}+{\omega_{\infty}}^{2} \right) {\omega_{\infty}}^{4}}{ \left( {\omega_{e}}^{2}-{\omega_{\infty}}^{2}
 \right) ^{3}{r}^{4}}}.
\end{equation}
\end{widetext}

\begin{widetext}
W only consider $\mathcal{K}dS$ at first order in $m$: 
\begin{equation}
\mathcal{K}dS=-{\frac { \left(  \left( 2\,r-3\,{\it M_0} \right) {\omega_{\infty}}^{
4}-3\,{\omega_{e}}^{2} \left( r+{\it M_0} \right) {\omega_{\infty}}^{2}+r{\omega_{e}}^{4} \right) \zeta\,\omega_{\infty}\,{
\it M_0}}{ \left( -{\omega_{e}}^{2}+{\omega_{\infty}}^{2}
 \right) ^{5/2}{r}^{3}}}
dr d\varphi+\mathcal{O}(M_0^{2}).
\end{equation}
\end{widetext}
On the other hand,  we have
\begin{equation}
\frac{d\sigma}{d\varphi}\bigg|_{C_{R}}=
n ( R ) \left( \frac { \zeta^2 R^2 } { f ( R ) } \right) ^ { 1 / 2 },
\end{equation}
thus, we show that it goes to $\zeta$:
\begin{equation}
\lim_{R\to\infty} \kappa_g\frac{d\sigma}{d\varphi}\bigg|_{C_R}=\zeta\,.
\end{equation}
For the limit of  $R\to\infty$, and using the straight light approximation $r=b/\sin\varphi$, the Gauss-Bonnet theorem becomes \cite{Crisnejo:2018uyn}:
\begin{equation}
\lim_{R\to\infty} \int^{\pi+\alpha}_0 \left[\kappa_g\frac{d\sigma}{d\varphi}\right]\bigg|_{C_R}d\varphi =\pi-\lim_{R\to\infty}\int^\pi_0\int^R_{\frac{b}{\sin\varphi}}\mathcal{K} dS.
\end{equation}
Hence, the deflection angle with the approximate expression for the weak field limits reads as:
\begin{equation}\label{eq:perf}
\alpha=4\,\pi\,\mu+\,{\frac {4{\it M_0}}{b}}+4\,{\frac {{\it M_0}\,{\omega_{e}}^{2}}{{
\omega_{\infty}}^{2}b}}
\end{equation}
These results show that the photon rays move in a medium of homogeneous plasma. It is noted that $\omega_e/\omega_\infty\to 0$, \eqref{eq:perf} reduces to \eqref{deflection}, and the effect of the plasma can be removed.

\section{Conclusion}
In this paper, we performed a comprehensive study of RBCS's deflection angle of photons in weak field approximation. To this end, we have used the Gauss-Bonnet theorem and a straight line approximation to obtain the deflection angle of light at leading order terms. Then, we calculated the deflection angle of light by RBCS in a plasma medium up to the first order with the approximate expression for the weak deflection. For both cases, the cosmic string parameter $\mu$ and the mass term $M_0$ increase the deflection angle. After neglecting the plasma effects, $\omega_e/\omega_\infty\to 0$, \eqref{eq:perf} reduces to \eqref{deflection}. The deflection angle using the Gauss-Bonnet theorem is calculated by integrating over a domain outside the impact parameter, which shows that gravitational lensing is a global effect and is a powerful tool to research the nature of black hole singularities.

\section*{Acknowledgement}
This work is supported by Comisi{\'o}n Nacional de
Ciencias y Tecnolog{\'i}a of Chile (CONICYT) through FONDECYT Grant N{$%
\mathrm{o}$} 3170035 (A. {\"O}.).

\end{document}